\documentstyle[aps,prl,preprint]{revtex}

\begin{document}
\draft
\title{Strong Effects of Weak Localization in Charge Density Wave/Normal Metal
Hybrids}
\author{Mark I. Visscher, B. Rejaei, and Gerrit E.W. Bauer}
\address{Theoretical Physics Group, Department of Applied Physics and Delft Institute%
\\
of Microelectronics and Submicron Technology, Delft University of\\
Technology, \\
Lorentzweg 1, 2628 CJ Delft, The Netherlands}
\date{\today}
\maketitle

\begin{abstract}
Collective transport through a multichannel disordered conductor in contact
with charge-density-wave electrodes is theoretically investigated. The
statistical distribution function of the threshold potential for
charge-density wave sliding is calculated by random matrix theory. In the
diffusive regime weak localization has a strong effect on the sliding motion.
\end{abstract}


Wave (Anderson) localization is an important concept of the quantum theory
of electron transport. It is well established by now that a random disorder
potential may cause bound states with an exponential decay governed by the
localization length. When this localization length is large compared to
other length scales in the system the corrections to semiclassical transport
are small and the term ``weak localization'' is appropriate. Weak
localization is well understood for normal Fermi liquids. However,
localization in heterostructures like superconducting/normal metal (S/N)\
or\ Josephson\ (S/N/S) junctions is still actively investigated,
experimentally and theoretically. For example, the electron-hole coherence
induced in a normal metal by a superconducting reservoir significantly
modifies the weak localization correction to the electron transport through
the hybrid structure \cite{Beenakker}.

A charge-density wave (CDW) is a collective state similar to BCS
superconductors (S) \cite{Gruener}. For instance, CDW reservoirs induce a
proximity effect in normal metals (N) analogous to that in N/S junctions
when the interface is parallel to the CD wave fronts (normal to the chains).
In such a geometry the phase coupling through a defect-free CDW/N/CDW
junction supports a sliding CDW mode \cite{Visscher}. A process analogous to
Andreev scattering at an N/CDW interface\ has been predicted \cite{Rejaei}.
Advances in thin film growth and patterning of CDW compounds \cite{Zant0}
enables the study of CDW transport in various geometries on a mesoscopic
scale \cite{Latyshev0,Nina}. Quantum oscillations have been identified in
the magnetoresistance of the CDW compound {\rm NbSe}$_{3}$ in the presence
of columnar defects \cite{Latyshev}. Electrical contacts which inject the
current parallel to the CDW chains allow detailed studies of the subgap
conductance \cite{Sinch}.

Here we report characteristic effects of weak localization on CDW
sliding through a disordered conductor. In contrast to the extensive
literature on semiconductor- or superconductor hybrid structures \cite
{Beenakker} this topic has to the best of our knowledge not yet been
investigated. The statistical properties of diffuse conductors including the
weak localization correction to the average conductance are well described
by random matrix theory \cite{Beenakker1}. We calculate the statistical
properties of the pinning energy in the presence and absence of
time-reversal symmetry by averaging over random transfer matrices. We find
that the pinning strength is much more sensitive to the
localization-enhanced backward scattering than the normal conductance. Weak
localization is destroyed when time reversal-symmetry is broken by a
magnetic field \cite{Mello1}, leading to a substantial negative
magnetoresistance.

We consider a disordered metal with length $L$ and width $W$ sandwiched
between two CDW reservoirs consisting of $N$ half-infinite chains (Fig. \ref
{Figure}). The CDW state is characterized by a complex order parameter $%
\Delta \exp (i\chi ),$ where the amplitude $2\Delta $ is the gap in the
electronic spectrum and its phase $\chi $ denotes the position of the CDW
relative to the crystal lattice. The amplitude of the order parameter $%
\Delta $ is taken to be equal on all chains and constant up to the
interfaces. In the limit of large correlation lengths $\xi =\hbar
v_{F}/\Delta \gg L$, the Coulomb interaction between the chains and the
phase coherence over the region allows us to define a single CDW phase $\chi 
$ for the whole structure. For simplicity the material parameters of the CDW
and the normal metal are taken to be matched at the interface. This
corresponds to a normal metallic island fabricated by ion damage \cite
{Latyshev} or selective abrasion \cite{Zant} of the CDW material. Mismatch
between a CDW and {\em e.g.} a true 2D or 3D metal or imperfect interfaces
cause additional scattering which, in principle, can be incorporated into
our formalism.

Many semi-conducting CDW materials are characterized by a nearly
one-dimensional electron band structure. On the other hand, there are also
CDW materials with substantial transverse dispersion, like {\rm NbSe}$_{3}$.
For the latter systems and in the diffusive $\left( L\gg l\right) $ and
quasi-one-dimensional $\left( L\gg W\right) $ limits an electron flux\ from
one chain is distributed evenly among the outgoing chains. These two
``anisotropic'' and ``isotropic'' limiting cases (see Fig. \ref{Figure}) are
treated below separately. Interpolation between these two extremes
interpolation seems reasonable.

In a Wannier site representation with respect to individual chains,
equilibrium properties of CDW's can be described by the retarded and
advanced quasi-classical Green functions $(g_{\alpha \beta }{\bf )}%
_{ij}^{R,A}(x,\epsilon ),$ where $\alpha ,\beta =\{R,L\}$ and $%
i,j=\{1,...,N\}$ \cite{Artemenko}. The subscripts $\{R,L\}$ refer to the
right and left moving electrons on the linearized branches of the electronic
spectrum, and $i,j$ are the chain indices. The $2N\times 2N$ matrix Green
functions Left (-) and Right (+) of the scattering region are related by the
boundary condition 
\begin{equation}
{\bf M}^{\dagger }{\bf g}_{+}={\bf g}_{-}{\bf M}^{\dagger },  \label{boundy}
\end{equation}
where ${\bf M}$ is the $2N\times 2N$ transfer matrix of the disordered
region which satisfies the condition ${\bf M}^{\dagger }{\bf \bbox{\Sigma}}%
_{3}{\bf M\bbox{\Sigma}}_{3}={\bf 1}$ in order to ensure current
conservation, and ${\bf M}^{\ast }={\bf \bbox{\Sigma}}_{1}{\bf M\bbox{\Sigma}%
}_{1}$ in the presence of time-reversal symmetry. Here ${\bf \bbox{\Sigma}}%
_{1}$ and ${\bf \bbox{\Sigma}}_{3}$ are defined as 
\begin{equation}
{\bf \bbox{\Sigma}}_{1}=\left( 
\begin{array}{cc}
{\bf 0} & {\bf 1} \\ 
{\bf 1} & {\bf 0}
\end{array}
\right) ,\quad {\bf \bbox{\Sigma}}_{3}=\left( 
\begin{array}{cc}
{\bf 1} & {\bf 0} \\ 
{\bf 0} & -{\bf 1}
\end{array}
\right) .
\end{equation}
Additional scattering at the interfaces mentioned above can be incorporated
by substituting ${\bf M}\rightarrow {\bf M}_{R}{\bf MM}_{L}$, where ${\bf M}%
_{R,L}$ are the transfer matrices of the right and left interfaces,
respectively. The transfer matrix can be decomposed as \cite{Mello1} 
\begin{equation}
{\bf M}=\left( 
\begin{array}{ll}
\bbox{\alpha}^{(1)} & {\bf 0} \\ 
{\bf 0} & \bbox{\alpha}^{(3)}
\end{array}
\right) \left( 
\begin{array}{cc}
({\bf 1}+\bbox{\lambda})^{1/2} & \bbox{\lambda}^{1/2} \\ 
\bbox{\lambda}^{1/2} & ({\bf 1}+\bbox{\lambda})^{1/2}
\end{array}
\right) \left( 
\begin{array}{ll}
\bbox{\alpha}^{(2)} & {\bf 0} \\ 
{\bf 0} & \bbox{\alpha}^{(4)}
\end{array}
\right) ,  \label{mdecom}
\end{equation}
where $\bbox{\alpha}^{(i)}$ with $i=\{1,2,3,4\}$ are $N\times N$ unitary
matrices and $\bbox{\lambda}$ is a real diagonal matrix with elements $%
0<\{\lambda _{1,}...,\lambda _{N}\}<\infty $, frequently referred to as
eigenparameters of ${\bf M}$. In the presence of time-reversal symmetry $%
\bbox{\alpha}^{(1)\ast }=\bbox{\alpha}^{(3)}$ and $\bbox{\alpha}^{(2)\ast }=%
\bbox{\alpha}^{(4)}.$

Backscattering represented by the non-diagonal elements of ${\bf M}$ pins
the CDW condensate, $i.e.$ no current flows in linear response. By forcing
the system out of the equilibrium position $\chi _{0}$ a chemical potential
difference between left and right side arises. The threshold potential $\mu
_{T}$ at which the condensate starts to slide corresponds to the maximum
value of the chemical potential difference as a function of the condensate
position $\chi $: 
\begin{equation}
\mu _{T}/2\equiv \max \left[ \mu _{+}(\chi )-\mu _{-}(\chi )\right] .
\label{thresdef}
\end{equation}
$\left[ \mu _{+}-\mu _{-}\right] (\chi )$ can be identified as the periodic
pinning potential. In terms of the retarded and advanced Green functions the
chemical potentials $\mu _{\pm }$ of left and right CDW's read 
\begin{equation}
\mu _{\pm }=-\frac{1}{8N}\int d\epsilon \,\tanh (\frac{\epsilon }{2k_{B}T})\,%
{\rm Tr\,}{\bf \bbox{\Sigma}}_{3}\left( {\bf g}_{\pm }^{R}-{\bf g}_{\pm
}^{A}\right) .  \label{chempot}
\end{equation}
The pinning potential is expressed in terms of the transfer matrix elements
via the Green functions ${\bf g}_{\pm }^{R,A}$, which can be obtained from
their equation of motion \cite{Artemenko} and the boundary condition Eq. (%
\ref{boundy}). In the energy integral of Eq. (\ref{chempot}), the continuum
contribution to the energy integration dominates the bound states
contribution when $\epsilon _{F}\gg \Delta $. In this limit the bound states
on the island may be disregarded and we arrive via Eq. (\ref{thresdef}) at
the simple expression: 
\begin{equation}
\mu _{T}=\frac{\Delta }{N\gamma }\left| {\rm Tr}({\bf r}^{\prime }-{\bf r}%
^{\dagger })\right| ,  \label{thresr}
\end{equation}
where ${\bf r}^{\prime }=\bbox{\alpha}^{(1)}\frac{\bbox{\lambda}}{\bbox{1}+%
\bbox{\lambda}}\bbox{\alpha}^{(3)\dagger }$ and ${\bf r}=-\bbox{\alpha}^{(4)}%
\frac{\bbox{\lambda}}{\bbox{1}+\bbox{\lambda}}\bbox{\alpha}^{(2)\dagger }$
are the $(N\times N)$ reflection matrices of the disordered region and $%
\gamma \ll 1$ is the dimensionless electron-phonon coupling constant \cite
{note}. The threshold potential (\ref{thresr}) is proportional to a factor
containing CDW material parameters and a ``form factor'' depending on
transfer matrix elements of the disordered region. In contrast to the
reflection probability ${\rm Tr\,}{\bf r}^{\dagger }{\bf r}$, $\mu _{T}$ is
not invariant under a transformation ${\bf r}\rightarrow {\bf urv}$, where $%
{\bf u},{\bf v}$ are arbitrary unitary matrices. This is a direct
consequence of the broken translational invariance of CDW's in the presence
of defects. Naturally, the threshold is positive definite. By extending it
symmetrically to negative threshold values the distribution function can be
constructed from the averages of all even moments.

The theory of random matrices is first employed for the isotropic limit
described above. The ensemble average of a quantity $f({\bf M})$ of a
collection of random conductors with length $L$ is defined as 
\begin{equation}
\left\langle f\right\rangle =\int f({\bf M})\,p_{L}(\bbox{\lambda})\,d\mu (%
{\bf M})  \label{mataverage}
\end{equation}
where $d\mu ({\bf M})$ is the invariant or Haar measure and $p_{L}(%
\bbox{\lambda})$ is the isotropic probability density \cite{Mello1}. In
terms of the parameterization in Eq. (\ref{mdecom}), the measure is
explicitly given by 
\begin{equation}
d\mu ({\bf M})=J_{\beta }(\bbox{\lambda})d^{N}{\bf \lambda }%
\prod_{i=1}^{2\beta }d\mu (\bbox{\alpha}^{(i)}),
\end{equation}
where the Jacobian 
\begin{equation}
J_{\beta }(\bbox{\lambda})=\prod_{a<b}|\lambda _{a}-\lambda _{b}|^{\beta },
\end{equation}
and $d\mu (\bbox{\alpha}^{(i)})$ are the invariant measures of the unitary
group$.$ The symmetry index $\beta =1$ describes the orthogonal ensemble in
the presence of time-reversal symmetry, and the unitary ensemble $\beta =2$
implies broken time-reversal symmetry. The isotropic distribution function $%
P(\bbox{\lambda}{\bf ,}L)=p_{L}(\bbox{\lambda})J(\bbox{\lambda})$ satisfies
an $N$-dimensional Fokker-Planck equation with the mean-free path $l$ as
single parameter \cite{Mello1}. From the calculated average in Eq. (\ref
{mataverage}) it is possible to extract the distribution ${\cal P}(\mu _{T})$
of the threshold indirectly via

\begin{equation}
\left\langle f\right\rangle =\int_{0}^{\infty }d\mu _{T}\,f(\mu _{T}){\cal P}%
(\mu _{T}).  \label{average}
\end{equation}
In the following $\left\langle {}\right\rangle _{\alpha }$ and $\left\langle
{}\right\rangle _{\lambda }$ denote the average over unitary matrices and
eigenparameters, respectively.

We start by considering the ensemble averages of the $2n$-th$\ $powers of
Eq. (\ref{thresr}). All even moments $\left\langle \mu
_{T}^{2n}\right\rangle ^{(\beta )}$ can be generated from the quantity 
\begin{equation}
\left\langle e^{\eta {\rm Tr}({\bf r}^{\prime }-{\bf r}^{\dagger }){\rm Tr}(%
{\bf r}^{\prime \dagger }-{\bf r})}\right\rangle ^{(\beta )},
\label{exponent}
\end{equation}
by a Taylor expansion in $\eta .$ It is convenient to write this as a
Gaussian integral over the generating variables $\varphi $ and $\varphi
^{\ast }$%
\begin{equation}
e^{\eta {\rm Tr}({\bf r}^{\prime }-{\bf r}^{\dagger }){\rm Tr}({\bf r}%
^{\prime \dagger }-{\bf r})}=\frac{1}{2\pi i\eta }\int d\varphi \,d\varphi
^{\ast }e^{-|\varphi |^{2}/\eta +\varphi {\rm Tr}({\bf r}^{\prime \dagger }-%
{\bf r})+\varphi ^{\ast }{\rm Tr}({\bf r}^{\prime }-{\bf r}^{\dagger })}.
\end{equation}
In this form we can average ${\bf r}$ and ${\bf r}^{\prime }$ separately
over the unitary matrices, since the number of normal and Hermitian
conjugated matrices must be equal in order to give non-zero contributions to
the average \cite{Mello3}. Equation (\ref{exponent}) may therefore be
rewritten as 
\begin{equation}
\left\langle \frac{1}{\eta }\int_{0}^{\infty }d|\varphi |^{2}e^{-\left|
\varphi \right| ^{2}/\eta }\left[ \sum_{n=0}^{\infty }\frac{1}{(n!)^{2}}%
|\varphi |^{2n}\left\langle \left( {\rm Tr\,}{\bf r}^{\dagger }{\rm Tr\,}%
{\bf r}\right) ^{n}\right\rangle _{\alpha }\right] ^{2}\right\rangle
_{\lambda }^{(\beta )}.  \label{expansion}
\end{equation}
The first order term in the summation $(n=1)$ can be evaluated exactly 
\begin{equation}
\left\langle {\rm Tr\,}{\bf r}{\rm Tr\,}{\bf r}^{\dagger }\right\rangle
^{(\beta )}=\frac{2}{\beta N+2-\beta }\left\langle R\right\rangle _{\lambda
}^{(\beta )},  \label{firstorder}
\end{equation}
where we used $R={\rm Tr\,}{\bf r}^{\dagger }{\bf r=}{\rm Tr\,}\bbox{\lambda}%
(\bbox{1}+\bbox{\lambda})^{-1}.$ This result agrees with the microscopic
calculations of Dorokhov \cite{Dorokhov} for the statistical distribution of
the reflection matrix of a thin wire with Gaussian white noise potentials
and inter-chain hopping. From this relation we learn that the threshold
potential is determined by the (diagonal) reflection probabilities $%
\left\langle |r_{ii}|^{2}\right\rangle ^{(\beta )}$ at each chain. The $%
\beta $ dependent prefactor reflects the enhancement of backscattering by a
factor $2$, which is suppressed when time-reversal symmetry is broken \cite
{Mello1}. It is instructive to compare this result with the reflectance of
the disordered region $\left\langle {\rm Tr\,}{\bf r}^{\dagger }{\bf r}%
\right\rangle ^{(\beta )}=\left\langle R\right\rangle _{\lambda }^{(\beta )}$
which differs only by the absence of the prefactor. The localization
correction to the normal resistance recognized by its $\beta $-dependence is
called {\it weak }because it vanishes to lowest order in $N$. In contrast,
the pinning due to CD modulations induced by the reservoirs is {\it strongly}
affected. In the limit $N\gg 1$ the (squared) pinning strength is {\em %
doubled} $\left\langle \mu _{T}^{2}\right\rangle ^{(1)}=2\left\langle \mu
_{T}^{2}\right\rangle ^{(2)}$ by the weak-localization enhanced
backscattering at each chain.

For the $n$-th power $\left\langle \left( {\rm Tr\,}{\bf r}^{\dagger }{\rm %
Tr\,}{\bf r}\right) ^{n}\right\rangle _{\alpha }$ we found analytical
expressions only for large $N$. In the Gaussian limit \cite{Brouwer} 
\[
\left\langle \left( {\rm Tr\,}{\bf r}^{\dagger }{\rm Tr\,}{\bf r}\right)
^{n}\right\rangle _{\alpha }=n!\left( \frac{2}{\beta }\right) ^{n}\left( 
\frac{\left\langle R\right\rangle _{\lambda }^{(\beta )}}{N}\right) ^{n}, 
\]
\begin{equation}
\left\langle e^{\eta {\rm Tr}({\bf r}^{\prime }-{\bf r}^{\dagger }){\rm Tr}(%
{\bf r}^{\prime \dagger }-{\bf r})}\right\rangle ^{(\beta )}=\left[ 1-\frac{%
4\eta }{\beta }\left( \frac{s}{s+1}\right) \right] ^{-1}
\end{equation}
to leading order ${\cal O}(N^{0})$. Here, the average over the
eigenparameters $\bbox{\lambda}$ has been performed to leading order in $N$
using the classical solution $\left\langle R^{n}\right\rangle _{\lambda
}^{(\beta )}=\left( Ns/\left( s+1\right) \right) ^{n}$ with $s=L/l$ \cite
{Beenakker1}. Comparing the coefficients with powers of $\eta $ on both
sides, we finally arrive at a Wigner-like distribution function 
\begin{equation}
{\cal P}_{I}^{(\beta )}(\mu _{T})=\frac{2\beta \mu _{T}}{\left\langle \mu
_{T}^{2}\right\rangle }e^{-\beta \mu _{T}^{2}/\left\langle \mu
_{T}^{2}\right\rangle },
\end{equation}
for $0<\mu _{T}<\infty ,$ and $\left\langle \mu _{T}^{2}\right\rangle
=2\left( \Delta /N\gamma \right) ^{2}\left( s/s+1\right) .$ The mean value
and the standard deviation $\sigma $ are easily calculated as $\left\langle
\mu _{T}\right\rangle ^{(\beta )}=\Delta /N\gamma \sqrt{\pi s/\beta (s+1)}$
and $\sigma =\sqrt{3}\approx 1.732.$ By breaking time-reversal symmetry a
magnetic field destroys the weak-localization enhanced backscattering, thus
reducing the average threshold potential by a factor of $1/\sqrt{2}\approx
0.71$ even though the normal resistance remains unchanged!

We now examine the anisotropic limit in which interchain scattering is
absent. Time-reversal symmetry cannot be broken by a magnetic field so we
can restrict the discussion to the $\beta =1$ ensemble. The distribution of
the unitary matrices is\ not random but sharply centered around the unit
matrix. Therefore we take the unitary matrices to be diagonal. For a weak
impurity $\lambda \ll 1$, located on a single chain $N=1$ we obtain in the
ballistic limit $\left( s\ll 1\right) :$%
\begin{equation}
{\cal P}_{A}(\mu _{T})=\frac{2}{\sqrt{\pi \left\langle \mu
_{T}^{2}\right\rangle }}e^{-\mu _{T}^{2}/\left\langle \mu
_{T}^{2}\right\rangle }\qquad N=1,  \label{dis1}
\end{equation}
with $\left\langle \mu _{T}^{2}\right\rangle =\left( 2\Delta /\gamma \right)
^{2}s.$ The distribution of pinning energies of a defect localized to a
single chain satisfies a ``half'' Gaussian distribution, with an average $%
\left\langle \mu _{T}\right\rangle =\left( 2\Delta /\gamma \right) \sqrt{%
s/\pi }.$ In the large-$N$ expansion the probability for zero threshold
vanishes according to 
\begin{equation}
{\cal P}_{A}(\mu _{T})=\frac{2\mu _{T}}{\left\langle \mu
_{T}^{2}\right\rangle }e^{-\mu _{T}^{2}/\left\langle \mu
_{T}^{2}\right\rangle }\qquad N\gg 1,
\end{equation}
where $\left\langle \mu _{T}^{2}\right\rangle =\frac{2}{N}\left( \frac{%
\Delta }{\gamma }\right) ^{2}\left( \frac{s}{s+1}\right) $ for arbitrary
lengths $s$.

A negative magnetoresistance has been observed in ${\rm NbSe}_{3}$ in the
presence of columnar defects, which is caused by the reduced threshold
potential in high magnetic fields \cite{Latyshev}. In semiconductor CDW
materials as{\rm \ Rb}$_{0.3}${\rm Mn0}$_{3}$ or {\rm TaS}$_{3}$ the effect
has to our knowledge not been reported. These observations are consistent
with our findings since the anisotropy of the normal state conductance of 
{\rm NbSe}$_{3}$ is low, in contrast to the fully gapped, strongly
anisotropic CDW materials.

Our results also support previous work related to the periodicity of the
Aharonov-Bohm oscillations \cite{Visscher1}. As shown here, the average
threshold potential of a multichannel disordered conductor is sensitive to
weak-localization trajectories. In the weak pinning limit, the effective
threshold potential of many uncorrelated columnar defects as modelled by
metallic islands is given by its averaged value. Therefore, we can argue
that the latter is periodic in $h/2e$, since this component reflects the
contribution from time-reversed paths. Finally, we point out that in the
presence of many uncorrelated metallic islands the distribution of
thresholds smears out the threshold singularity in the current-voltage
characteristics, and determines the functional behavior of the non-linear
conductance.

We conclude by summarizing our results. We present expressions for the
threshold potential for collective CDW current through a multichannel
disordered conductor in terms of its reflection matrices. By means of random
matrix theory we calculate the statistical distribution of the pinning
strength with and without time-reversal symmetry. We find that localization
which is weak on the normal conductance has a strong effect on the threshold
potential. The breaking of time-reversal symmetry by a magnetic field
reduces the threshold field, depending on the isotropy of the disordered
conductor. The magnitude of the reduction ranges from zero in the
anisotropic limit to $1/\sqrt{2}$ in the isotropic limit.

It is a pleasure to thank Carlo Beenakker for valuable discussions. This
work is part of the research program for the ``Stichting voor Fundamenteel
Onderzoek der Materie'' (FOM). We acknowledge support from the NEDO joint
research program (NTDP-98).


\begin{references}
\bibitem{note}  For simplicity we have taken ${\bf M}$ to be energy
independent. ${\bf M}$ is then evaluated for particles at the Fermi surface.
This is a good approximation when the dwelling time of the particles on the
island is small compared to $\hbar /\Delta .$
\end{references}
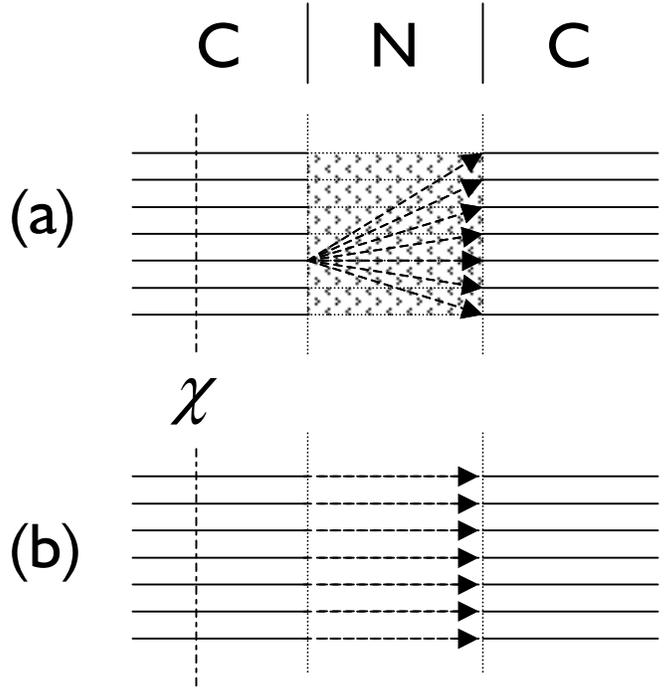
\end{document}